\documentclass[twocolumn,english,aps,pre,superscriptaddress]{revtex4}
\usepackage{amsmath}
\usepackage{graphicx}
\usepackage{amssymb}

\makeatletter
\@ifundefined{textcolor}{}
{%
 \definecolor{BLACK}{gray}{0}
 \definecolor{WHITE}{gray}{1}
 \definecolor{RED}{rgb}{1,0,0}
 \definecolor{GREEN}{rgb}{0,1,0}
 \definecolor{BLUE}{rgb}{0,0,1}
 \definecolor{CYAN}{cmyk}{1,0,0,0}
 \definecolor{MAGENTA}{cmyk}{0,1,0,0}
 \definecolor{YELLOW}{cmyk}{0,0,1,0}
 }

\makeatother

\usepackage{babel}

\begin{document}

\global\long\def\avg#1{\langle#1\rangle}

\global\long\def\p{\prime}

\global\long\def\dg{\dagger}

\global\long\def\ket#1{|#1\rangle}

\global\long\def\bra#1{\langle#1|}

\global\long\def\proj#1#2{|#1\rangle\langle#2|}

\global\long\def\inner#1#2{\langle#1|#2\rangle}

\global\long\def\tr{\mathrm{tr}}

\global\long\def\pd#1#2{\frac{\partial#1}{\partial#2}}

\global\long\def\spd#1#2{\frac{\partial^{2}#1}{\partial#2^{2}}}

\global\long\def\der#1#2{\frac{d#1}{d#2}}

\global\long\def\im{\imath}

\renewcommand{\onlinecite}[1]{\cite{#1}}

\title{Driving denaturation: Nanoscale thermal transport as a probe of DNA
melting}

\author{Kirill A. Velizhanin }

\email[]{kirill@lanl.gov}

\affiliation{Theoretical Division, Los Alamos National Laboratory, Los Alamos,
NM 87545}

\affiliation{CNLS, Los Alamos National Laboratory, Los Alamos, NM 87545}

\author{Chih-Chun Chien}

\affiliation{Theoretical Division, Los Alamos National Laboratory, Los Alamos,
NM 87545}

\author{Yonatan Dubi}

\thanks{Current address: School of Physics and Astronomy, Tel Aviv University,
Israel.}

\affiliation{Theoretical Division, Los Alamos National Laboratory, Los Alamos,
NM 87545}

\author{Michael Zwolak}

\email[]{mpz@lanl.gov}

\affiliation{Theoretical Division, Los Alamos National Laboratory, Los Alamos,
NM 87545}

\begin{abstract}
DNA denaturation has been a subject of intense study due to its relationship
to DNA transcription and its fundamental importance as a nonlinear
structural transition. Many aspects of this phenomenon, however, remain
poorly understood. Existing models fit quite well with experimental
results on the fraction of unbound base pairs versus temperature,
but yield incorrect results for other essential quantities, such as
the base pair fluctuation timescales. Here, we demonstrate that nanoscale
thermal transport can serve as a sensitive probe of the underlying
microscopic physics responsible for the dynamics of DNA denaturation.
Specifically, we show that the heat transport properties of DNA are
altered significantly as it denatures, and this alteration encodes
detailed information on the dynamics of thermal fluctuations and their
interaction along the strand. This finding allows for the discrimination
between models of DNA denaturation and will help shed new light on
the nonlinear vibrational dynamics of this important molecule. 
\end{abstract}
\maketitle
Besides being the {}``molecule of life'' \textendash{} or perhaps
because of it \textendash{} DNA lives at a unique position where its
double-stranded structure can unravel into two single strands, i.e.,
it denaturates or {}``melts'', via changes in conditions such as
temperature or ionic concentration \cite{Sinden94-1}. Local melting
of DNA due to thermal fluctuations, which can occur well below the
denaturation temperature, is thought to play a major role in the formation
of the transcription bubble \cite{Chen10-1}. The denaturation of
DNA proceeds via the thermal breakage (dissociation) of base pairs
and a nonlinear interaction between base pairs sharpens this transition
via cooperative effects in base pair unbinding \cite{Peyrard89-1,Dauxois93-1,Dauxois93-2}.
As a result, the accurate description of mechanisms of DNA denaturation
requires one to not only correctly account for local thermal fluctuations
but also how these fluctuations interact along the strand. These processes
are typically probed \emph{indirectly} through the measurement of
the DNA melting curve \textendash{} the fraction of unbound base pairs
versus temperature at equilibrium \textendash{} thus hindering the
understanding of the denaturation mechanisms. However, thermal fluctuations
and their interaction along the strand are the same processes that
occur in thermal transport, which, as we shall see, can give an independent
and more direct assessment of the proposed DNA denaturation mechanisms.

During the past decade, there has been tremendous progress in measuring
the thermal and electronic properties of molecular systems \cite{Dubi09-1,Heath09-1},
including a recent experiment on the heat conduction of inhomogeneous
DNA-Gold nanocomposites \cite{Kodama09-1}. However, except for few
studies relevant to double-stranded DNA \cite{Terraneo02-1,Peyrard06-1,Savin10-1},
no existing analysis of thermal transport takes into account DNA's
large structural fluctuations that eventually result in its denaturation.

\begin{figure}
\centering{}\includegraphics[width=7cm]{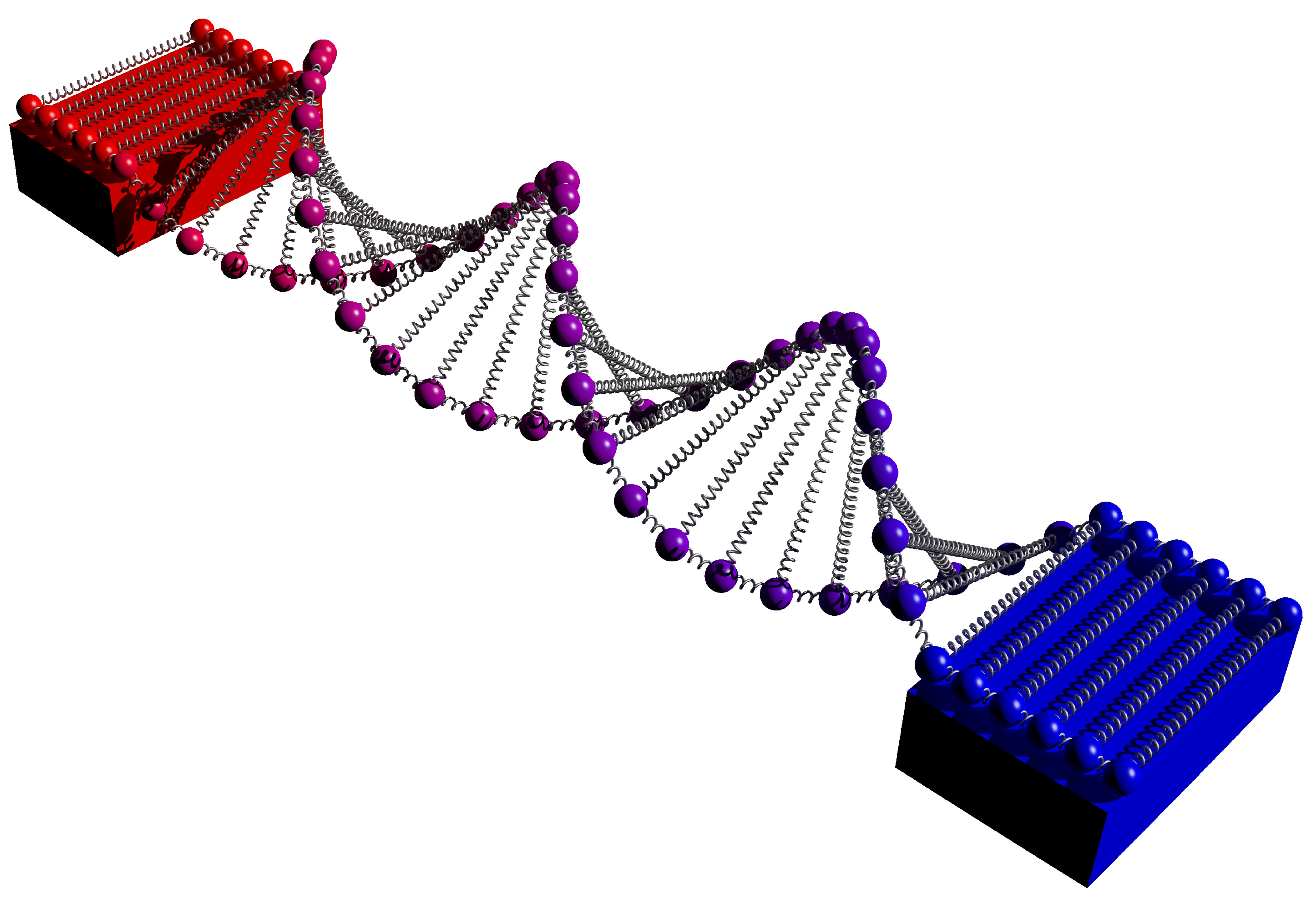} \caption{\label{fig:Schematic} Schematic of DNA between two heat reservoirs
that probe its structure via an energy current.}

\end{figure}

In this paper, the non-equilibrium behavior of single-coordinate nonlinear
models of DNA are analyzed as heat is driven through DNA via two thermal
reservoirs, as shown schematically in Fig.~\ref{fig:Schematic}.
The preeminent single-coordinate model for thermally driven denaturation
of DNA is the Peyrard-Bishop-Dauxois (PBD) model \cite{Peyrard89-1,Dauxois93-1,Dauxois93-2},
which is believed to capture the main physical processes behind DNA
denaturation \cite{Peyrard04-1}. By first deriving an analytic expression
for the thermal conductance, $\kappa$, we show that the PBD model
predicts a substantial jump in $\kappa$ upon DNA denaturation. That
is, the conductance of DNA is predicted to actually \emph{increase},
despite the fact that the DNA becomes disordered as it is broken into
two single strands. This change is a delicate balance of the model
parameters and the mechanisms that it represents. Indeed, we show
that another related model \cite{Joyeux05-1}, which also can describe
the statistical properties of DNA denaturation, gives qualitatively
different non-equilibrium behavior, predicting a drop in $\kappa$.
Thus, by driving DNA out of equilibrium, the basic physical aspects
of denaturation \textendash{} thermal fluctuations and their coupling
along the molecule \textendash{} can be probed.

We begin with a description of the PBD model \cite{Dauxois93-1}.
This model, as well as other single-coordinate models for DNA, reduces
the DNA strand to a one-dimensional, classical lattice with each base
pair represented by a single coordinate \textendash{} the collective
stretching of the hydrogen bonds in a Watson-Crick pair. The PBD Hamiltonian
takes on the form \begin{equation}
H=\sum_{n}\left[\frac{m\dot{y}_{n}^{2}}{2}+V\left(y_{n}\right)+W\left(y_{n},y_{n-1}\right)\right],\label{eq:H}\end{equation}
 where $y_{n}$ represents the stretching coordinate within the $n^{{\rm th}}$
base pair. An on-site Morse potential, $V(y_{n})=D(e^{-ay_{n}}-1)^{2}$,
describes the harmonic binding of the base pair at small $y_{n}$
with the angular frequency $\sqrt{2Da^{2}/m}$, and unbinding at $y_{n}\gg a$
with the finite dissociation energy $D$. The PBD nearest-neighbor
potential is given by \begin{equation}
W(y_{n},y_{n-1})=\frac{K}{2}(1+\rho e^{-\alpha(y_{n}+y_{n-1})})(y_{n}-y_{n-1})^{2}.\label{eq:PBD}\end{equation}

The PBD model has been quite successful in characterizing statistical
properties of DNA near the denaturation transition \cite{Campa98-1,Weber09-1}.
However, since it was not derived from a microscopic Hamiltonian,
one has to validate the model by comparison with experimental results.
In this regard, there is a considerable ambiguity in the results for
short DNA strands \cite{Campa98-1,Peyrard08-1,Sanrey09-1,Ares09-1}
as well as the timescales for the opening of bubbles at room temperature
\cite{Peyrard08-1,Peyrard09-1}. Thermal transport, as we show below,
can determine some of the main features single-coordinate models should
display (beyond equilibrium denaturation curves) and help settle the
above discrepancies.

We start our analysis by examining the behavior of the PBD model in
the low ($L$) and high ($H$) temperature limits, where it becomes
harmonic with the effective Hamiltonian \begin{equation}
H_{\mu}=\sum_{n}\left[\frac{m\dot{y}_{n}^{2}}{2}+D_{\mu}y_{n}^{2}+\frac{K_{\mu}}{2}\left(y_{n}-y_{n-1}\right)^{2}\right],\label{eq:HHL}\end{equation}
 with $\mu=L,\, H$. Numerical studies show that the full PBD model,
described by the Hamiltonian in Eq.~(\ref{eq:H}), undergoes a sharp
transition corresponding to the denaturation of DNA \cite{Dauxois93-2}.
Above the transition temperature, $T_{c}$, the base pairs become
unbound, i.e., the on-site potential approaches a constant leading
to $D_{H}=0$. The nearest-neighbor potential becomes harmonic with
$K_{H}=K$. In the low temperature limit, the strand will again have
harmonic behavior with the parameters of the effective Hamiltonian
$D_{L}=Da^{2}$ and $K_{L}=K\left(1+\rho\right)$ (the latter reflects
the larger nearest-neighbor coupling of the low temperature strand).
Even at low-temperatures, the nonlinearity of the potential is important,
but the harmonic limit gives the correct physical interpretation of
the numerical results on the full model we present later.

We introduce the concept of the thermal conductance ratio, $R$, as
the ratio of the conductance between the high and low temperature
phases $R\equiv\kappa_{H}/\kappa_{L}$. For Ohmic heat reservoirs
strongly coupled to the first and last site of the strand, it can
be calculated to be (see Appendix~\ref{app:Analytics} and Refs.~\cite{Casher71-1,Dhar01-1} for details)
\begin{equation}
R\approx\frac{2K_{H}D_{L}}{K_{L}^{2}},\label{eq:R}
\end{equation}
which uses the lowest order term in $K_{L}/D_{L}$ for $\kappa_{L}$
(see Eq.~(\ref{eq:tcond}) in Appendix~\ref{app:Analytics}). Since Eq.~(\ref{eq:R})
is for strong coupling, it represents an extreme value for $R$. For
the PBD model with the original parameters from Ref.~\cite{Dauxois93-1},
the corresponding ratio is $R_{PBD}\approx18$ \cite{ftn:parms}.
That is, the transition from low to high $T$ results in a substantial
jump in $\kappa$ due to the weakening of the on-site confining potential,
i.e., the base pair dissociation, as $T$ crosses $T_{c}$ from below.
The removal of the on-site potential softens the modes of the strand,
making them more efficient in the transport of heat. This is more
transparent if one examines the low temperature conductance, \begin{equation}
\kappa_{L}\propto\frac{K_{L}^{2}}{D_{L}}\propto\frac{K_{L}^{2}}{m\omega_{L}^{2}},\label{eq:kL}\end{equation}
 where $\omega_{L}$ is the on-site frequency and we have taken $K_{L}/D_{L}$
to be small. Equation~(\ref{eq:kL}) shows that the conductance increases
with decreasing frequency.

We note that the PBD model give a novel phenomenon compared to other
transitions. For example, in the ice-water transition at 273 K, the
conductance decreases, rather than increases, with a ratio of $R_{\mathrm{H_{2}O}}\approx0.25$
\cite{CRC}. This is mainly due to the breaking of crystal structure
across the transition that reduces the coupling between molecules.
Another example is single-crystal C$_{60}$, which exhibits a face-centered-cubic
to simple-cubic structural transition at 260 K \cite{Yu92-1}. The
ratio for this transition can be estimated from the experimental data
to be $R_{C_{60}}\approx0.75$. 

The thermal conductance ratio is thus important in identifying the
underlying mechanisms of a structural transition. The balance of the
softening of the lattice and the reduction of the nearest neighbor
coupling determines both the magnitude of the conductance change,
as well as whether it increases or decreases. Different proposals
for the nonlinear forms of the model will give different limiting
harmonic forms, changing the conductance behavior across the temperature
range not only quantitatively, but also qualitatively. For example,
another single-coordinate model of DNA is that of Joyeux and Buyukdagli
(JB) in Ref.~\cite{Joyeux05-1}. It uses the same on-site potential
but employs an alternative nearest-neighbor potential that, although
giving a very similar melting curve, results in a different high temperature
harmonic Hamiltonian (see Appendix~\ref{app:Analytics}). In the strong-coupling
limit, $R$ changes as \[
R_{PBD}\approx18\longrightarrow R_{JB}\approx0.017.\]
 That is, the PBD model gives qualitatively distinct behavior from
the JB model because the latter looses virtually all of its nearest
neighbor coupling at high temperature, i.e., it looks more like the
ice-to-water transition in character. This qualitative conclusion
does not change even when using different parameter sets that have
been developed \cite{ftn:parms1}.
Therefore, while they both agree well with some denaturation
experiments, the conductance ratio can be used to directly probe the
physics contained within their respective interaction potentials.
While temperatures much higher than the transition temperature are
beyond the validity of the models, the opposing predictions persist
within a range of temperatures, including around the denaturation
transition. In addition, having the ends of the DNA ``clamped''
together on each of the heat reservoirs -- one way to measure its
thermal transport properties -- will help the fluctuations described
by these single-coordinate models to persist to higher temperatures.
This strongly supports using the heat conduction as a test to determine
basic aspects of DNA denaturation that need to be represented in microscopic
models.

\begin{figure}
\centering{}\includegraphics[width=8cm]{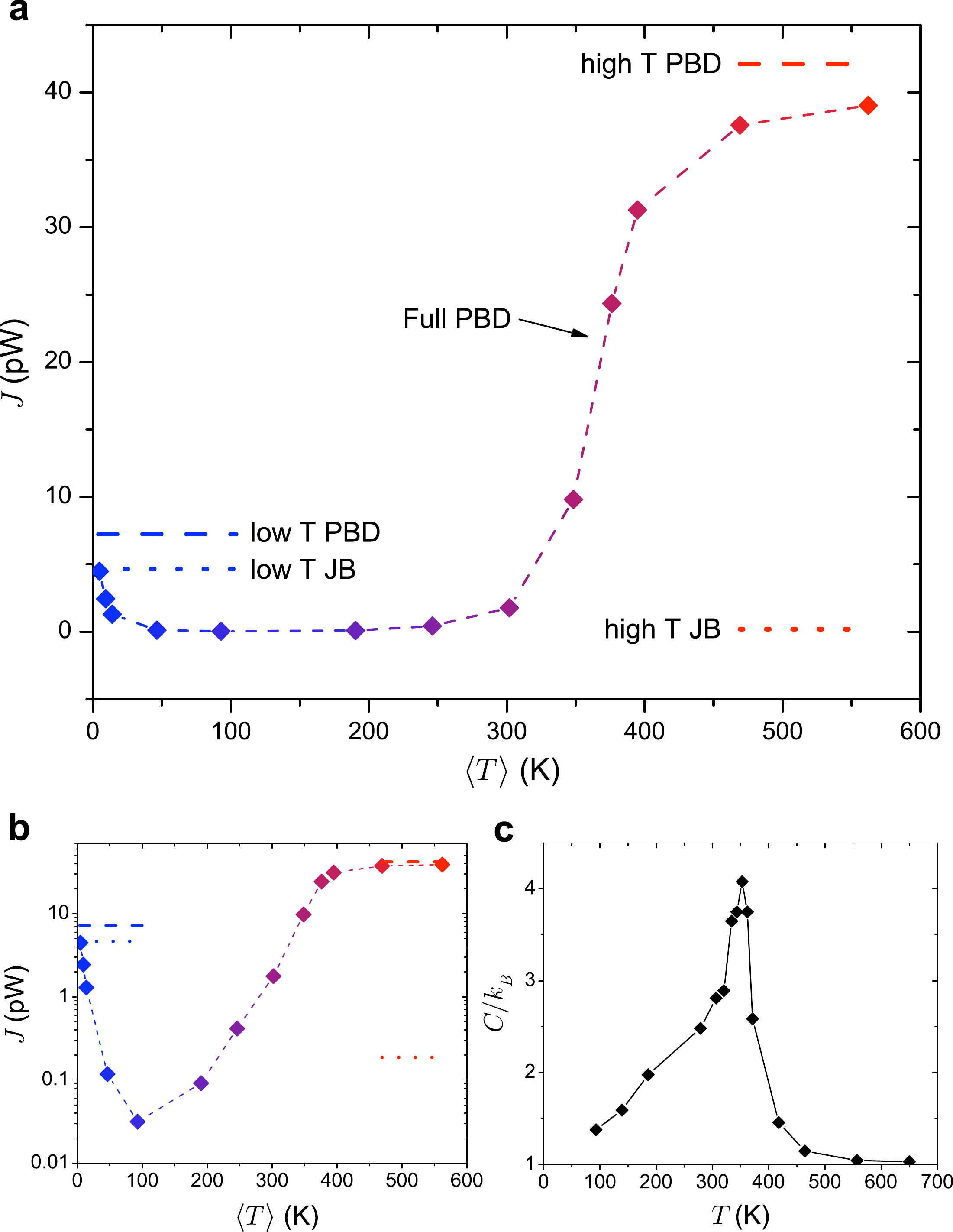} \caption{\label{fig:ThermalCurrent}Thermal transport across the denaturation
transition. (a) Thermal current through DNA as a function of average
temperature, $\langle T\rangle=\left(T_{L}+T_{H}\right)/2$, calculated
while keeping the temperature difference $\Delta T=T_{H}-T_{L}=9.3\,\mathrm{K}$
constant. The thermal current displays a steep rise as the temperature
crosses the transition temperature $T_{c}\simeq350\,\mathrm{K}$.
Low and high temperature simulations of the harmonic strands defined
by $H_{L}$ and $H_{H}$ are given by blue and red lines, respectively.
The numerical results on the full PBD model approach the harmonic
limits at sufficiently low and high temperatures. (b) Logarithmic
scale plot of the thermal current. The convergence to the harmonic
limit, as well as an non-monotonic dependence of the thermal current
on temperature at low temperatures and a sharp rise near the transition
are visible. The latter indicates that DNA may be suitable to function
as a molecular thermal switch. (c) Dependence of equilibrium heat
capacity, $C$, (normalized per length of the strand) on temperature
showing the transition at $T_{c}$ and the convergence to the harmonic
value $C/k_{B}=1$. The convergence to the harmonic value at high
temperature is much more rapid than a low temperature, indicating
that the anharmonicity of the Morse potential is still important even
at low temperature.}

\end{figure}

To gain further insight into the non-equilibrium behavior of DNA,
we solve for the full dynamics of the PBD model (Eq.~(\ref{eq:H})
with the appropriate stacking interaction) by numerically integrating
the equations of motion of the DNA between two Langevin reservoirs
(see Appendix~\ref{app:Numerics} for details on numerical simulations).
In Fig.~\ref{fig:ThermalCurrent}(a),
we plot the thermal current as a function of the average reservoir
temperature, while keeping the temperature difference constant at
a value significantly lower than the width of the phase transition
so that the entire strand is always in one phase. %

As seen Fig.~\ref{fig:ThermalCurrent}, the ability of the DNA to conduct heat increases
substantially after it becomes denaturated according to the PBD model,
thus confirming our analytical prediction.\textbf{ }Further, the conductivity
at $\avg T=300\,\mathrm{K}$ is 0.18~pW/K\AA{} \textendash{} smaller
than that in Ref.~\cite{Savin10-1}. The difference is due to different
interactions taken into account, which result in an absence of any
signatures of denaturation up to 350 K in Ref.~\cite{Savin10-1}.
Thus, their results do not incorporate the effect of strong nonlinearities
due to denaturation/local melting.\textbf{ }We have also calculated
the thermal current for purely harmonic strands with the parameters
corresponding to the different phases (as discussed below Eq.~(\ref{eq:HHL})).
The thermal conductance ratio for the harmonic limits is found numerically
to be $R_{PBD}\approx6$, which is lower than our analytical result
due to the more realistic reservoir implemented in our simulations
(see Appendix~\ref{app:Numerics}). Figure~\ref{fig:ThermalCurrent}(c)
plots the heat capacity $C$ (per base pair) as a function of temperature,
indicating the denaturation transition and the approach to harmonicity
(where $C/k_{B}$ is always unity).

The nonlinear model, however, gives much richer behavior. While at
high temperature there is convergence of the nonlinear model to its
harmonic limit, at low temperatures a non-monoticity appears, more
clearly seen in Fig.~\ref{fig:ThermalCurrent}(b) where we plot the
thermal current as a function of temperature on a logarithmic scale.
The non-monotonicity, reflected in the drop of $\kappa$ in the temperature
range $0-100$ K, is mainly due to phonon scattering off the anharmonic
on-site potential. This signature \textendash{} conductance versus
temperature \textendash{} indicates the strength of the nonlinearity
in DNA dynamics, which we will investigate further in a future contribution.
Although the single-coordinate models we study here were proposed
to study DNA behavior near the denaturation transition, such a nonlinear
signature should still appear at low temperature, implying that a
measurement of DNA thermal transport could also serve to define the
temperature range of validity of the PBD model. It also gives rise
to a three orders of magnitude increase of the thermal conductance,
indicating that this characteristic of structural phase transitions
can be useful in developing heattronic devices such as thermal switches
and thermal transistors \cite{Chien11-1}.

In the above discussion we have considered only homogeneous DNA strands.
It is clear that heterogeneity will give rise to interesting effects
in the conductance. Alternating sequences of AT and GC pairs, for
instance, will tend to suppress the low temperature conductance due
to the mismatch between parameters of base pair binding potentials,
while the high temperature conductance will remain roughly the same.
This will result in a substantial increase in $R$ \cite{Chien11-1},
which will also occur due to random or genomic DNA. Differences between
AT and GC pairs gives an additional experimental test of this class
of models \cite{Chien11-1}.

To conclude, we have investigated the out-of-equilibrium properties
of single-coordinate models of DNA in order to elucidate the nonlinear
nature of denaturation. The balance of lattice softening and reduced
coupling in different models leads to qualitatively distinct predictions
for the thermal conductance ratio \textendash{} the experimental measurement
of which will illuminate these fundamental aspects of DNA and thus
should shed light on the underlying mechanisms of DNA denaturation.
Due to the nature of the effect, experiments will be robust with respect
to heterogeneity, defects, and contacts, unlike the electrical counterpart
\cite{Diventra04-1}. Recent experiments on the heat conduction of
nanotubes \cite{Kim01-1} and DNA-Gold nanocomposites \cite{Kodama09-1}
give a potential route to developing the experimental setup. However,
it will be necessary to perform the experiment in conditions of high
humidity so the DNA retains its B-form \cite{Record81-1}. Such an
experiment could also be done on double- and single- stranded DNA
separately, which would remove uncertainties due to the lack of solvent.
Fluorescence experiments to detect local openings of the DNA strand
\cite{Bonnet98-1,Bonnet03-1} can further complement the thermal transport
measurements. The thermal conductance ratio may also be useful in
understanding other structural transitions, such as nanotube collapse.
Finally, such experiments will pave the road to creating molecular
thermal switches and open new avenues in understanding and controlling
thermal flow at the nanoscale. 
\begin{acknowledgments}
The authors would like to thank A. R. Bishop, K. Rasmussen and B.
Alexandrov for valuable discussions. This research is supported by
the U.S. Department of Energy through the LANL/LDRD Program. K.A.V.
also acknowledges the support provided by CNLS.
\end{acknowledgments}

\appendix

\section{Analytic derivation of the thermal conductance}\label{app:Analytics}

The thermal conductance of a classical harmonic lattice can be found
analytically. Our starting point is the procedure of Refs.~\cite{Casher71-1,Dhar01-1}.
We consider the limiting cases of the single-coordinate Hamiltonian, Eq.~(\ref{eq:H}),
for a lattice of length $N$. One already
saw that the low- (L) and high- (H) temperature limits can be approximated
by a harmonic Hamiltonian of the form
\begin{equation}
H_{\mu}=\sum_{n}\left[\frac{m\dot{y}_{n}^{2}}{2}+D_{\mu}y_{n}^{2}+\frac{K_{\mu}}{2}\left(y_{n}-y_{n-1}\right)^{2}\right],\end{equation}
where $\mu=L,\: H$.

To simplify the analysis, the lattice is coupled to two heat reservoirs
at the first and last sites, which gives the equations of motion \begin{align}
m\ddot{y}_{n} & =-2(D_{\mu}+K_{\mu})y_{n}+K_{\mu}(y_{n-1}+y_{n+1})\nonumber \\
 & +(\delta_{n,1}+\delta_{n,N})\left[\int_{-\infty}^{t}dt^{\prime}A(t-t^{\prime})y_{n}(t^{\prime})+\eta_{n}(t)\right].\end{align}
 We choose the spectrum of the dissipation to be ohmic, $A(\omega)=-i\gamma\omega$,
with coupling $\gamma$, and the noise to be white noise, $\langle\eta_{L/H}(\omega)\eta_{L/H}(\omega^{\prime})\rangle=4\pi T_{L/H}\gamma\delta(\omega+\omega^{\prime})$,
with $T_{L/H}$ the low and high reservoir temperatures. This form
for the reservoirs will satisfy the fluctuation-dissipation theorem.
The resulting equations of motion are \begin{align}
m\ddot{y}_{n} & =-2(D_{\mu}+K_{\mu})y_{n}+K_{\mu}(y_{n-1}+y_{n+1})\nonumber \\
 & +(\delta_{n,1}+\delta_{n,N})\left[-\gamma\dot{y}_{n}(t)+\eta_{n}(t)\right].\end{align}
 The solution for the coordinates has the form
\begin{equation}
y_{n}(t)=(1/2\pi)\int_{-\infty}^{\infty}d\omega\hat{Y}_{nm}^{-1}(\omega)\hat{\eta}_{m}(\omega)e^{i\omega t},
\end{equation}
where $\hat{\eta}$ is a vector of length $N$ with the first and
last components being $\eta_{L/H}(\omega)$ and the rest being zero.
It represents the coupling of the reservoirs to the ends of the lattice.
The $N\times N$ matrix $\hat{Y}=\hat{\phi}-\omega^{2}\hat{M}-\hat{A}$
encodes the solution. Here $\hat{\phi}_{nm}=2(D_{\mu}+K_{\mu})\delta_{n,m}-K_{\mu}\delta_{n,m+1}-K_{\mu}\delta_{n,m-1}$,
$\hat{M}_{ij}=m\delta_{i,j}$, and $\hat{A}_{11}=\hat{A}_{NN}=A(\omega)$
and $\hat{A}_{nm}=0$ otherwise.

The heat current flowing into the lattice is $J=\langle[\int_{-\infty}^{t}dt^{\prime}A(t-t^{\prime})y_{1}(t^{\prime})]\dot{y}_{1}(t)\rangle$,
where the average is over the noise. Setting $\gamma=\lambda m$,
the heat current becomes \begin{align}
J_{\mu} & =\frac{\Delta T\lambda^{2}m^{2}}{\pi}\int_{-\infty}^{\infty}d\omega\omega^{2}\{(\mathcal{D}_{1,N}-\lambda^{2}\omega^{2}m^{2}\mathcal{D}_{2,N-1})^{2}\nonumber \\
 & +\lambda^{2}\omega^{2}m^{2}(\mathcal{D}_{1,N-1}+\mathcal{D}_{2,N})^{2}\}^{-1}|C_{1,N}|^{2},\end{align}
 where $\Delta T=T_{H}-T_{R}$ is the temperature difference of the
reservoirs, $C_{1,N}$ is the cofactor of $\hat{Y}_{1,N}$, and $\mathcal{D}_{n,m}$
is the determinant of the submatrix of $(\hat{\phi}-\omega^{2}\hat{M})$
from the $n$-th row (column) to the $m$-th row (column). It follows
that $|C_{1,N}|^{2}=K_{\mu}^{2N-2}$ and $\mathcal{D}_{n,m}=K_{\mu}^{n-m+1}\mathcal{D}_{n,m}^{0}$.
The elements of $\mathcal{D}_{nm}^{0}$ are given by $\left(\begin{array}{cc}
\mathcal{D}_{1,N}^{0} & -\mathcal{D}_{1,N-1}^{0}\\
\mathcal{D}_{2,N}^{0} & -\mathcal{D}_{2,N-1}^{0}\end{array}\right)=\mathcal{T}^{N}$, where \begin{equation}
\mathcal{T}=\left(\begin{array}{cc}
2(1+D_{\mu}/K_{\mu})-(m/K_{\mu})\omega^{2} & -1\\
1 & 0\end{array}\right).\end{equation}
 We look for the eigenvalues of $\mathcal{T}$ of the form $\exp(\pm iq)$
with real $q$ because those eigenvalues correspond to propagating
modes. This requirement imposes that $2\cos(q)=2(1+D_{\mu}/K_{\mu})-(m/K_{\mu})\omega^{2}$.
After changing variables from $\omega$ to $q$ that satisfy this
constraint, the final expression (for an infinite lattice ($N\rightarrow\infty$))
is \begin{equation}
\frac{J_{\mu}}{\Delta T}=\frac{\gamma}{2\pi m}\int_{0}^{2\pi}dq\frac{\sin^{2}(q)}{1+\frac{2\gamma^{2}}{mK_{\mu}}\left[1+\frac{D_{\mu}}{K_{\mu}}-\cos(q)\right]}.\end{equation}
 This gives, explicitly for the low and high temperature thermal conductance
$\kappa_{\mu}=J_{\mu}/\Delta T$, \begin{equation}
\kappa_{\mu}=\frac{k_{B}mK_{\mu}^{2}}{4\gamma^{3}}\left[1+\frac{2\gamma^{2}}{mK_{\mu}}+\frac{2\gamma^{2}D_{\mu}}{mK_{\mu}^{2}}-\mathcal{B_{\mu}}\right],\label{eq:fullkappa}\end{equation}
 with \begin{equation}
\mathcal{B}_{\mu}=\sqrt{1+\frac{4\gamma^{2}}{mK_{\mu}}+\frac{4\gamma^{2}D_{\mu}}{mK_{\mu}^{2}}+\frac{8\gamma^{4}D_{\mu}}{m^{2}K_{\mu}^{3}}+\frac{4\gamma^{4}D_{\mu}^{2}}{m^{2}K_{\mu}^{4}}}.\end{equation}
 With these expressions one can explicitly find the thermal conductance
ratio $R$. We have verified that for reservoirs contacted to a single
site on each end, the thermal conductance from our numerical simulations
agree with our analytic formula to within $10-15$\%, which we attribute
to finite size effects in the numerical simulations.

We can take various limiting forms of these equations. If we define
the prefactor as $\tilde{\kappa}_{\mu}$ and a dimensionless reservoir
coupling as \begin{equation}
\gamma_{\mu}=\frac{\gamma}{\sqrt{mK_{\mu}}},\end{equation}
the expressions for the conductance become
\begin{equation}
\kappa_{\mu}=\tilde{\kappa}_{\mu}\left[1+2\gamma_{\mu}^{2}+2\gamma_{\mu}^{2}\frac{D_{\mu}}{K_{\mu}}-\mathcal{B_{\mu}}\right],
\label{eq:tcond}
\end{equation}
with
\begin{equation}
\mathcal{B}_{\mu}=\sqrt{1+4\gamma_{\mu}^{2}+4\gamma_{\mu}^{2}\frac{D_{\mu}}{K_{\mu}}+8\gamma_{\mu}^{4}\frac{D_{\mu}}{K_{\mu}}+4\gamma_{\mu}^{4}\left(\frac{D_{\mu}}{K_{\mu}}\right)^{2}}.
\end{equation}
The appropriate limiting forms for our case are the following. When
the high temperature harmonic limit has no onsite potential, then
the heat conductance becomes \begin{equation}
\kappa_{H}=\tilde{\kappa}_{H}\left[1+2\gamma_{H}^{2}-\sqrt{1+4\gamma_{H}^{2}}\right].\end{equation}
 For a low temperature limit that has a much greater onsite term than
the nearest neighbor coupling, i.e., $K_{L}/D_{L}\ll1$, the heat
conductance becomes \begin{equation}
\kappa_{L}\approx\frac{\tilde{\kappa}_{L}\gamma_{L}^{2}K_{L}}{D_{L}},\end{equation}
 which also assumes that the dimensionless coupling to the reservoirs
is $\gamma_{L}\geq1$. For strong coupling to the reservoirs, the
ratio becomes \begin{equation}
R\approx\frac{2K_{H}D_{L}}{K_{L}^{2}}.\end{equation}
 This is the analytic expression we use within the article. The strong
coupling limit gives the extreme value of $R$.

Now we show that the characteristic frequency in the PBD model is
lowered as $T$ crosses $T_{c}$ from below. For the low temperature
Hamiltonian, the corresponding equation of motion is \begin{equation}
m\ddot{y}_{n}=-\{2Da^{2}y_{n}+K(1+\rho)[(y_{n}-y_{n-1})+(y_{n}-y_{n+1})]\}.\end{equation}
 From the ansatz $y_{n}=y_{n}^{0}e^{i\omega t-ikn}$ , one obtains
the phonon spectrum as \begin{equation}
m\omega^{2}=2Da^{2}+2K(1+\rho)[1-\cos(k)].\end{equation}
 Thus, the frequency band of phonons is $\sqrt{2Da^{2}/m}\le\omega\le\sqrt{[2Da^{2}+4K(1+\rho)]/m}$.
For the high temperature Hamiltonian, the equation of motion is \begin{equation}
m\ddot{y}_{n}=-K[(y_{n}-y_{n-1})+(y_{n}-y_{n+1})]\end{equation}
 and the phonon spectrum is \begin{equation}
m\omega^{2}=2K[1-\cos(k)].\end{equation}
 The frequency band of phonons is $0\le\omega\le\sqrt{4K/m}$. The
two limiting Hamiltonians are both harmonic and the characteristic
frequency is indeed lowered, and similar considerations apply for
other models. In the low temperature limit, the onsite potential stiffens
the DNA compared to the high temperature limit, which results in the
raising of the phonon spectrum of the low temperature limit compared
to the high temperature. In the latter, as well, the nearest neighbor
coupling drops from $K\left(1+\rho\right)$ to $K$ shrinking the
bandwidth. This trade-off is responsible for the change in thermal
conductance across the transition. If the drop in nearest neighbor
coupling is small, then the softening will dominate, and the heat
conductance will increase because these softened modes can conduct
heat more effectively.

\section{Numerical Simulation Details}\label{app:Numerics}

To study the dynamics of the DNA out of equilibrium we solve numerically
the Langevin equation, which describes the dynamics of a Hamiltonian
system in the presence of thermal baths. The Langevin equation is
given by \begin{equation}
m\ddot{y}_{n}=-\frac{\partial W}{\partial y_{n}}-\frac{\partial V}{\partial y_{n}}-\Gamma_{n}\dot{y}_{n}+f(t),\end{equation}
where $W(y_{n})$ and $V(y_{n})$ are the potentials described in
Eq.~(\ref{eq:H}). The DNA strand is split into three regions,
the two ends, each of length $l$, serve as the Langevin thermal reservoirs
at temperatures $T_{L}$ and $T_{H}$. This means that the friction
term $\Gamma_{n}$ only operates for $n$ within the thermal reservoirs.
The fluctuating term $f(t)$ is Gaussian white noise which obeys the
fluctuation-dissipation relation $\langle f(t)f(t')\rangle=2\Gamma_{n}k_{B}T_{L(H)}\delta(t-t')$
for the low and high temperature reservoirs, respectively.

The middle region of the length $M$ is the free DNA strand, which
is driven out of equilibrium by the Langevin reservoirs when $T_{L}\neq T_{H}$.
The parameters for the simulations are $M=60$ and $l=20$, and the
PBD model parameters are $D=0.04\,\mathrm{eV}$, $a=4.47\,\textrm{\AA}^{-1}$,
$K=0.04\,\mathrm{eV/\textrm{\AA}^{2}}$ , $m=300\,\mathrm{u}$, $\rho=0.5$,
and $\alpha=0.358\,\textrm{\AA}^{-1}$ \cite{Dauxois93-1}. The strand
is homogeneous DNA except for the random forces and friction in the
Langevin regions. The equations of motion are integrated with the
fourth-order Runge-Kutta method. The temperature profile of the strand
is evaluated by defining the local temperature at site $n$ as $k_{B}T_{n}^{loc}=m\langle\dot{y}_{n}^{2}\rangle$,
where the average is over time. The local heat current is given by
$J_{n}=-\left\langle \dot{y}_{n}\frac{\partial W(y_{n},y_{n-1})}{\partial y_{n}}\right\rangle $.
The simulations are performed long enough to allow the system to reach
its steady state.

The simulations of the heat capacity were performed by connecting
Langevin reservoirs of identical temperature to every site in the
chain and evaluating the average energy per site by $E=\left\langle \frac{m\dot{y}_{n}^{2}}{2}+V(y_{n})+W(y_{n},y_{n-1})\right\rangle $,
where the averaging is performed over time and all the sites of the
chain. Then, the dependence of average energy per site on temperature,
$E(T)$, obtained by scanning through a set of temperature points,
was numerically differentiated to yield the heat capacity.

We also performed simulations on the JB model \cite{Joyeux05-1},
which has the same onsite potential as the PBD model and uses the
nearest neighbor potential \begin{equation}
W(y_{n},y_{n-1})=K_{b}(y_{n}-y_{n-1})^{2}+\frac{\Delta H}{2}(1-e^{-b(y_{n}-y_{n-1})^{2}}).\label{eq:JB}\end{equation}
 The parameters in this interaction are $K_{b}=10^{-5}\,\mathrm{eV/\textrm{\AA}^{2}}$,
$\Delta H=0.44\,\mathrm{eV}$, and $b=0.1\,\mathrm{\textrm{\AA}}^{-2}$.
Thus, the low temperature limit is $K_{L}^{JB}=2K_{b}+b\Delta H$,
with $D_{L}^{JB}=D_{L}$. The coupling $K_{L}^{JB}=0.044\,\mathrm{eV/\textrm{\AA}^{2}}$
is not substantially different from the PBD model, which has $K_{L}=0.06\,\mathrm{eV/\textrm{\AA}^{2}}$.
However, numerical simulations indicate that in the high temperature
limit the heat conductance converges to the conductance of a harmonic
Hamiltonian with nearest neighbor interaction $K_{b}(y_{n}-y_{n-1})^{2}$.
In other words, the quantity $(y_{n}-y_{n-1})^{2}$ is sufficiently
large to suppress the second term in Eq.~(\ref{eq:JB}). Thus, the
high temperature effective nearest neighbor coupling is $K_{H}^{JB}=2K_{b}$,
which is substantially smaller than the low temperature coupling,
and also substantially smaller than the high temperature coupling
of the PBD model.


\begin{thebibliography}{34}
\bibitem[{ftn({\natexlab{a}})}]{ftn:parms}
\bibinfo{note}{We use the original parameters for the PBD model from
  Ref.~\cite{Dauxois93-1} since there are still discrepancies in what
  parameters should be used (see, e.g.,
  Refs.~\cite{Weber09-1,Theodorakopoulos10-1,Theodorakopoulos11-1}). There is a
  large positive $R$ in the PBD model when changing physically relevant
  parameters, making it a robust feature of the model. Thermal transport
  measurements, as well, will help settle what parameters actually describe
  DNA.}
\end{thebibliography}

\end{document}